\documentclass{jetpl}
\twocolumn

\lat

\title{Bound on induced gravitational wave background \\ from primordial black holes}

\rtitle{Bound on induced gravitational wave background\ldots}

\sodtitle{Bound on induced gravitational wave background from primordial black holes}

\author{E.\,V.\,Bugaev,
P.\,A.\,Klimai\/\thanks{e-mail: pklimai@gmail.com}}

\rauthor{E.\,V.\,Bugaev, P.\,A.\,Klimai}

\sodauthor{Bugaev, Klimai}

\address{Institute for Nuclear Research, Russian Academy of
Sciences, 60th October Anniversary Prospect 7a, 117312 Moscow,
Russia}

\abstract{
The today's energy density of the induced (second order) gravitational wave background
in the frequency region $\sim 10^{-3} - 10^3$ Hz is constrained using the existing limits on
primordial black hole production in the early Universe. It is shown, in particular, that at
frequencies near $\sim 40$ Hz (which is the region explored by LIGO detector), the value of the
induced part of $\Omega_{GW}$ cannot exceed $(1-3)\times 10^{-7}$. The spread of values
of the bound is caused by the uncertainty in parameters of the gravitational collapse
of black holes.
}


\begin{document}

\maketitle

\section{Introduction}

It is well known now that gravitational waves (GWs) can be effectively generated by
density perturbations during the radiation dominated era.
Tensor and scalar perturbations are decoupled at the first order, but it is not so
in higher orders of cosmological perturbation theory. Namely, the primordial
density perturbations and the associated scalar metric perturbations generate
a cosmological background of GWs at second order through a coupling of modes
\cite{Matarrese:1993zf, Matarrese:1997ay, Carbone:2004iv}. In particular, a second order
contribution to the tensor mode, $h_{ij}^{(2)}$, depends quadratically on the first order
scalar metric perturbation, i.e., the observed scalar spectrum sources the generation of
secondary tensor modes. By other words, the stochastic spectrum of second order GWs is induced
by the first order scalar perturbations. Calculations of $\Omega_{GW}$ at second order and
discussions on perspectives of measurements of the second order GWs are contained in works
\cite{Mollerach:2003nq, Ananda:2006af, Baumann:2007zm, Saito:2008jc, Bugaev:2009zh}.

It is natural to conjecture that the detection of GWs from primordial density perturbations
on small scales (not directly probed by observations) could be used to constrain
overdensities on these scales. However, at the present time, gravitational wave background (GWB)
is not yet detected. So, on the contrary, one can constrain GWB using existing limits on
amplitudes of primordial density perturbations. Such limits are available, in particular,
from studies of primordial black hole (PBH) production at the beginning of radiation era.

It is generally known that PBHs form from the
density perturbations, induced by quantum vacuum fluctuations during inflationary expansion.
The details of the PBH formation had been studied in \cite{Carr:1975qj, Khlopov:1980mg}, the
astrophysical and cosmological constraints on the PBH density had been obtained in many
subsequent works (see, e.g., the recent review \cite{Khlopov:2008qy}).
The detailed constraints on amplitudes of density perturbation spectrum had been reviewed in
recent works \cite{Bugaev:2008gw, Josan:2009qn}.

Having the bounds on the primordial density spectrum, it is possible to determine the corresponding
bounds on amplitudes of produced GWB.

The plan of the paper is as follows. In Sec. 2 we present the basic relations connecting
the primordial power spectrum of density perturbations and induced GWB. The main result
of the paper - the PBH bound on the energy density of GWB - is shown in Fig. \ref{fig-gw-limit}.
In Sec. 3 the mini-review of theoretical estimates for $\Omega_{GW}$ is given.
Sec. 4 contains the conclusions.

\section{Derivation of the PBH bound}
\label{sec2}
The energy density of GWs per logarithmic interval of $k$ in units of the critical density is given by
\begin{equation} \label{OmegaGW}
\Omega_{\rm gw}(k, \tau) =
\frac{1}{12} \left( \frac{k}{a(\tau) H(\tau)} \right)^2 {\cal P}_h(k, \tau).
\end{equation}
The power spectrum ${\cal P}_h$ is defined by the standard expression
\begin{equation}
\langle h_{\bf k}(\tau) h_{\bf k'}(\tau) \rangle = \frac{1}{2} \frac{2\pi^2}{k^3}
\delta^3({\bf k}+{\bf k'}) {\cal P}_h(k, \tau),
\end{equation}
where $h_{\bf k}(\tau)$ is the Fourier component of the tensor metric perturbation.
For scalar-induced GWs, ${\cal P}_h(k, \tau)$ is obtained from the
formula \cite{Baumann:2007zm}
\begin{equation} \label{Phktau}
{\cal P}_h(k, \tau) = \int\limits_0^\infty d\tilde k \int \limits_{-1}^{1} d\mu  \;
{\cal P}_\Psi(|{\bf k-\tilde k|}) {\cal P}_\Psi (\tilde k) {\cal F}(k,\tilde k,\mu, \tau),
\end{equation}
where ${\cal P}_\Psi(k)$ is the power spectrum of the Bardeen potential, and
${\cal F}$ is the function determined by the transfer functions of the Bardeen
potentials and Green functions of the evolution equation for the GW amplitude,
\begin{equation} \label{h-evol}
h_{\bf k}'' + 2 {\cal H} h_{\bf k}' + k^2 h_{\bf k} = S({\bf k}, \tau).
\end{equation}
The source term in this equation is given by
\begin{equation} \label{Sk}
S({\bf k}, \tau) = \int d^3\tilde {\bf k} \; \tilde k^2(1-\mu^2) \; f({\bf k},\tilde {\bf k}, \tau)
\Psi_{ {\bf k}-\tilde {\bf k}} \Psi_{ \tilde {\bf k} },
\end{equation}
where $f$ can be explicitly expressed through the Bardeen potential's transfer functions
\cite{Baumann:2007zm}. It is seen from Eqs. (\ref{Phktau}, \ref{Sk}) that both the source term
and the power spectrum of induced background of second order
GWs depend quadratically on the first-order scalar perturbation $\Psi_{ {\bf k} }$.

It is also seen from the Eq. (\ref{Phktau}) that for scalar-induced GWs, the single
mode in scalar spectrum does not correspond to the only
one mode in ${\cal P}_h$. For example, for the $\delta$-function-like spectrum
${\cal P}_{\cal R}(k) \sim \delta(k-k_0)$, the GW spectrum turns out to be continuous and stretching from $0$
to $2k_0$ \cite{Ananda:2006af}. However, the order of magnitude of wave numbers of induced GWs
is typically the same as of scalar perturbations \cite{Bugaev:2009zh}, so, the GW's typical wave number
that will be generated from perturbations entering horizon at its mass scale $M_h$
can be estimated by
\begin{eqnarray} \label{kkeq}
k = k_{eq} \left( \frac{M_h}{M_{eq}} \right)^{-1/2} \left( \frac{g_*}{g_{* eq}} \right)^{-1/12}
\approx \\
\approx 2 \times 10^{23} (M_h[{\rm g}])^{-1/2} \;\; {\rm Mpc}^{-1} \nonumber,
\end{eqnarray}
where in the last equality we have adopted that $g_{* eq} \approx 3$, $g_* \approx 100$,
$M_{eq} = 1.3 \times 10^{49} {\rm g} \cdot (\Omega_m h^2)^{-2} \approx 8 \times 10^{50} {\rm g}$.
The connection between $f$ and $k$ for GW is
\begin{equation}
f=\frac{ck}{2\pi} = 1.54\times 10^{-15 } \left( \frac{k}{{\rm Mpc}^{-1}} \right) {\rm Hz}.
\label{fck}
\end{equation}

If PBHs form from a scalar spectrum of
perturbations at a horizon mass scale $M_h$, the typical
PBH mass will be of order of $M_h$ (see, e.g., \cite{Bugaev:2008gw}), so (\ref{kkeq})
relates the typical PBH mass with the characteristic $k$ of second order GWs produced.

We assume in this work, that the power spectrum ${\cal P}_{\cal R}(k)$ can have
a peak at some wave number $k_0$. Such peaks can arise in several types of
inflationary models (see, e.g., \cite{Bugaev:2008bi} and references therein)
and lead to PBH production, so that the peak's
parameters (such as height and width) can be constrained from PBH nonobservation.
It is convenient to use some kind of parametrization to model the realistic peaked power
spectrum of finite width, e.g.,
\begin{equation}
\label{PRparam} %
\lg {\cal P}_{\cal R} (k) = B + (\lg {\cal P}_{\cal R}^0 - B)
\exp \Big[-\frac{(\lg k/k_0)^2}{2 \Sigma^2} \Big],
\end{equation}
where $B \approx -8.6$, ${\cal P}_{\cal R}^0$ and $\Sigma$ characterize the height
and width of the peak, $k_0$ is the position of its maximum. Parameters of such
a distribution have been constrained previously \cite{Bugaev:2008gw} from non-observation
of PBHs and products of their Hawking evaporation.

The frequency region $\sim 10^3 - 10^{-3}\;$Hz which we explore in this work corresponds to horizon and PBH mass
region of $\sim 10^{11} - 10^{23}\;$g. For a rather wide peak ($\Sigma \gtrsim 1$), all constraints on ${\cal P}_{\cal R}^0$
for this region are in the
range $0.01 - 0.04$, depending on $k$ and the parameter $\delta_c$, which is a density contrast
threshold for PBH production (see \cite{Bugaev:2008gw} for details).

Particularly, it turns out that for mass range $\sim 10^{13}-10^{17}$g, the best constraints are obtained
from the requirement that the diffuse gamma ray
flux produced by PBHs does not exceed the observed extragalactic one.
The condition that the fraction of the energy density of the universe contained in PBHs, $\Omega_{\rm PBH}$,
does not exceed the one for non-baryonic dark matter, $\Omega_{\rm nbm} \approx 0.2$,
gives the most stringent limit for $M_{BH} \gtrsim 10^{17}\;$g
(this constraint exists for black holes with initial mass $M_{BH} > M_* \approx 5\times 10^{14}$ g, i.e., ones that did
not evaporate up to the present time).
For $M_{BH}>10^{13}\;$g we use constraints on ${\cal P}_{\cal R}^0$ obtained in \cite{Bugaev:2008gw}
using the above two conditions.

In the region $M_{BH} \sim 10^{11} - 10^{13}\;$g several types of constraints coexist, including
ones considering PBH influence on the photodissociation of deuterium \cite{Lindley1980, Clancy:2003zd},
CMB distortions \cite{Tashiro:2008sf}
and electron antineutrino background flux \cite{Bugaev:2008gw}.
It turns out that the constraint from deuterium photodissociation is the strongest one,
so for this mass region we use the constraints on ${\cal P}_{\cal R}$ of \cite{Josan:2009qn},
where limits derived in \cite{Clancy:2003zd} were updated.
One should note that an idea of constraining of PBH production rate from an influence of PBH evaporations
on the light element abundances had been proposed in pioneering paper \cite{Zeldovich1977}.
CMB distortion constraints obtained in \cite{Tashiro:2008sf} are basically about the same order of magnitude,
and limits from considering the neutrino background are somewhat weaker.
For even smaller PBH masses, other useful limits exist, e.g., the constraints from hadron injection by PBHs
during nucleosynthesis \cite{Zeldovich1977} (see the review \cite{Josan:2009qn}), but we do not explore this region here.

Numerical integration of (\ref{Phktau}) with the input given by (\ref{PRparam}) shows
that for realistic peak with the finite width (particularly, $\Sigma \gtrsim 1$), the
good estimate for maximum value of $\Omega_{GW}(k)$ is
\begin{equation}
\label{OmegaApprox2}
\Omega_{GW}^{max}(\tau_0) \approx \Omega_{GW}(k_0, \tau_0) \cong 0.002 \left( \frac{g_{* eq}}{g_*} \right)^{1/3}
\cdot ({\cal P}_{\cal R}^0)^2 ,
\end{equation}
i.e., maximum value does not depend on $\Sigma$ and is approached at the scale $k_0$ (the
particular shape of $\Omega_{GW}(k)$, of course, depends on $\Sigma$ - see analysis in
the work \cite{Bugaev:2009zh}). Since the PBH constraints on  ${\cal P}_{\cal R}^0$ weakly
depend on $\Sigma$, formula (\ref{OmegaApprox2}) can be used to obtain the limits on maximum possible
$\Omega_{GW}(k)$ for second order GWs.

\begin{figure}
\includegraphics[trim = 0 0 0 0, width=0.48 \textwidth]{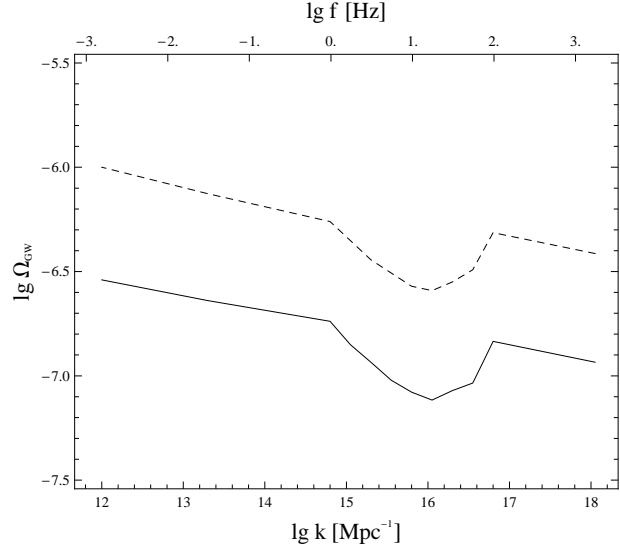} %
\caption{Figure 1. Constraints on the values of $\Omega_{GW}$ for induced GW background, calculated
using formula (\ref{OmegaApprox2}) and existing PBH constraints on ${\cal P}_{\cal R}(k)$.
The forbidden values of $\Omega_{GW}$ lie above the curves shown in the figure.
Upper curve: constraints obtained using
the assumption of critical collapse, with threshold value of density contrast $\delta_c=0.45$;
lower curve: the case of standard collapse with $\delta_c=1/3$.} \label{fig-gw-limit}
\end{figure}

Fig. \ref{fig-gw-limit} shows the resulting constraints on $\Omega_{GW}(k)$ calculated using
known PBH limits on the scalar spectrum. The two curves correspond to the constraints
obtained for two different gravitational collapse models (standard and critical)
that assume different values of $\delta_c$. Since for critical collapse larger $\delta_c$
was used, larger values of $\Omega_{GW}(k)$ are possible in this case.

It should be noted that in the case of a very narrow, $\delta$-function like, spectrum
of ${\cal P}_{\cal R}(k)$, the constraints on ${\cal P}_{\cal R}^0$ used by us, of course, do not
hold anymore, and the formula (\ref{OmegaApprox2}) can not be applied. However,
the analysis shows \cite{Saito:2008jc} that in this case
maximal possible GW energy density is $\Omega_{GW}^{max} \sim 10^{-7} - 10^{-8}$ in the
range of frequencies which we consider ($\sim$ mHz - kHz), i.e., the values are smaller than
bounds of Fig. \ref{fig-gw-limit}. So the presented bounds are the actual upper limits on
$\Omega_{GW}(k)$.

Note that the constraints shown in Fig. \ref{fig-gw-limit} cover both the regions of $f$
that are currently probed by ground-based experiments LIGO and Virgo (best sensitivity at
$f\sim 100$ Hz) and the ones that will be accessible to future space-based experiments
LISA ($f \sim$ mHz) and BBO/DECIGO ($f\sim 0.1$ Hz).

\section{Theoretical estimates for the stochastic GWB}
\label{sec3}

\paragraph{Standard inflationary model.}
It has been derived in \cite{Starobinsky:1979ty} that de Sitter expansion phase in the early universe
generates the stochastic GWB with scale invariant power spectrum through the mechanism \cite{Grishchuk:1974ny}
of amplification of quantum vacuum
fluctuations. So, inflation is one of the most promising sources of GWB
\cite{Rubakov:1982df, Fabbri:1983us, Abbott:1984fp, Starobinsky:1985ww}.
The prediction for it is given by \cite{Starobinsky:1979ty, Allen:1999xw}
\begin{equation}
\Omega_{GW}^{inf}(k) h^2 \cong \Omega_r h^2 \frac{16}{9} \left( \frac{V_{inf} }{m_{Pl}^4} \right)
\left( \frac{g_0}{g_*} \right)^{1/3}.
\end{equation}
Here, $\Omega_r h^2 \approx 4 \times 10^{-5}$ is the total energy density fraction of radiation at present
time, $V_{inf}^{1/4}$ is the energy scale of inflation. From WMAP observations,
$V_{inf}^{1/4}<3.4 \times 10^{16}$ GeV, and  $\Omega_{GW}^{inf} h^2 < 2 \times 10^{-15}$
for $f> 10^{-16}$ Hz \cite{Smith:2005mm}.

\paragraph{GWB from preheating.}  The peak value of the background density is estimated by the
formula \cite{Khlebnikov:1997di, Easther:2006gt}
\begin{equation}
\Omega_{GW}^{pr}(k_{peak}) h^2 = \Omega_r h^2 \frac{\mu^2}{g^2 m_{Pl}^2}
\left( \frac{g_0}{g_*} \right)^{1/3}.
\end{equation}
It is assumed here that the inflationary potential at the end of inflation is given by
\begin{equation}
V_{inf}(\phi, \chi)  = \frac{1}{2} \mu^2 \phi^2 + \frac{1}{2} g^2 \phi^2 \chi^2,
\end{equation}
i.e., $\mu$ is the effective mass \cite{Easther:2006gt} and the frequency of an
oscillation of the field $\phi$, $g$ is the coupling constant. Energy scale of inflation
is
$V_{inf}^{1/4} \sim \sqrt{ \mu m_{Pl} }$,
the Hubble function is
$H_e = m_{Pl}^{-1} \sqrt{V_{inf}} \sim \mu$,
and the peak frequency (the predicted spectrum has a definite maximum) is
\begin{equation}
f_{peak} \approx 6\times 10^{10} \sqrt{ \frac{H_e}{M_{P}} } {\rm Hz} \approx
6\times 10^{10} \sqrt{ \frac{\mu}{M_{P}} } {\rm Hz}.
\end{equation}

If $\mu$ is very small, $\mu\sim 10^{-18} m_{Pl}$, then $f_{peak} \approx 200\;$Hz,
i.e., the peak is situated in the region studied by the LIGO detector.
In such a case, however, the coupling constant $g^2$ is extremely small, of
order of $10^{-30}$. Note that in standard models of chaotic inflation the parameter $\mu$
is not free, $\mu\sim 10^{-6} m_{Pl}$, and the peak frequency is $\sim 10^8\;$Hz.

The predictions of hybrid inflation models are more optimistic \cite{Dufaux:2007pt,
Dufaux:2008dn, GarciaBellido:2007af}. It was shown, in particular, that GWB can be relatively
large in the LIGO/BBO region. For the case $g^2/\lambda \ll 1$ the prediction of
\cite{Dufaux:2008dn} is
\begin{equation}
f_{peak} \approx \frac{g}{\sqrt{\lambda}} \lambda^{1/4} \; 10^{10.25} {\rm Hz},
\end{equation}
\begin{equation} \label{eq17}
h^2 \Omega_{GW} \ll 10^{-8.1} \left( \frac{\lambda}{g^2} \right)^{0.1}.
\end{equation}
Here, $\lambda$ and $g$ are parameters of the inflationary potential:
$\lambda$ is the Higgs self-coupling, $g$ is the Higgs-inflaton
coupling. For instance, for $\lambda\sim 0.1$, $f_{peak}\lesssim 10^3\;$Hz,
$g^2\lesssim 10^{-15}$ one has $h^2 \Omega_{GW} \lesssim 10^{-6}$.

In supersymmetric inflation models GWs can appear also as a result of an explosive decay
of flat direction condensates \cite{Dufaux:2009wn}. Typical frequencies of these GWs are in Hz-kHz range, and the
amplitude can be as high as $h^2 \Omega_{GW} \sim 10^{-6}$ \cite{Dufaux}.

\paragraph{First-order phase transitions in early Universe.}
The peak frequency is given by the formula \cite{Grojean:2006bp}
\begin{equation}
f_{peak} \approx 6\times 10^{-3} {\rm mHz} \left( \frac{g_*}{100} \right) ^ {1/6}
\left( \frac{T_*}{100\; {\rm GeV} } \right) \frac{f_*}{H_*}.
\end{equation}
Here, $H_*$ is the Hubble function at the time of GW production, $f_*$ is the characteristic
frequency of GWs, $T_*$ is the temperature at the time of the phase transition.
Typically, $f_*/H_* \sim 10^2$ \cite{Grojean:2006bp}, and $f_{peak}\sim 10^{-3}$ Hz if
$T_* \sim 100$ GeV and $f_{peak}\sim 10^{2}$ Hz if $T_* \sim 10^7$ GeV.

The energy density at the peak frequency is \cite{Grojean:2006bp}
\begin{equation}
\Omega_{GW}^{tr}(f_{peak}) h^2 \approx 10^{-5} \left( \frac{100}{g_*} \right)^{1/3}
\left( \frac{H_*}{f_*} \right)^2,
\end{equation}
i.e., $\Omega_{GW}^{tr}(f_{peak}) \sim 10^{-9}$.

\paragraph{Cosmic strings.} The estimate of the energy density is given
by the formula \cite{Hogan:2006we} (in high frequency limit, $f \gg 10^{-4}$ Hz)
\begin{equation}
\Omega_{GW}^{str} \sim 10^{-8} \left( \frac{G\mu}{10^{-9}} \right)^{1/2}
\left( \frac{\gamma}{50} \right)^{1/2}
\left( \frac{\alpha}{0.1} \right)^{1/2}.
\end{equation}
Here, $G\mu$ is the string tension, $\gamma$ is the radiation efficiency,
$\alpha$ is the initial loop size as a fraction of the Hubble radius.
Pulsar timing data constrain $\Omega_{GW}^{str}(f)$ at $f\sim 10^{-8}$ Hz,
and, according to this constraint \cite{DePies:2007bm}, one has,
at high frequency tail,
\begin{equation}
\Omega_{GW}^{str}  \lesssim 10^{-8}.
\end{equation}

\paragraph{Pre-big bang scenario.} In string cosmology (which is the theoretical base of the
pre-big bang (PBB) scenario \cite{Gasperini:2002bn}), the spectrum is growing with frequency, and the maximum value
is determined by the value of the parameter $M_s/M_P$ (where $M_s$ is the fundamental string mass).
According to the analysis of \cite{Gasperini:1999av}, the maximum estimate of the energy density is
\begin{equation}
\Omega_{GW}^{pbb, max}  < \Omega_\gamma \left( \frac{M_s}{M_P} \right)^2 \lesssim 10^{-6}.
\end{equation}
This maximum value is reached at $f\sim 100$ GHz, i.e., in the region which is inaccessible for an
experiment (at present time, at least). However, nonminimal variants of the scenario are
theoretically possible \cite{Gasperini:1999av}, in which the peak is shifted to lower frequencies
without changing the peak amplitude (e.g., to $f\sim 10^2$ Hz).

\paragraph{Brane-world models.} In these models, in which our world is a brane embedded in a higher
dimensional space, new geometrical degrees of freedom can be introduced (e.g., the position
of the brane). Excitations of these
degrees of freedom can lead, in principle, to detectable GW radiation. Naturally, the characteristic
frequencies depend on the size of the extra dimensions \cite{Hogan:2000is}.

\paragraph{GWB from PBH evaporations.} GWs are necessarily produced during PBH evaporations in early
universe. It is shown in \cite{BisnovatyiKogan:2004bk, Anantua:2008am} that the corresponding
stochastic GWB can be substantial, depending on the parameters of PBH formation models. Unfortunately,
typical frequencies of GWs are rather high ($10^{12}\;$Hz or more).

\section{Conclusions}
\label{sec4}

We have presented in the previous section that the typical predictions of the energy
density of GWB (characterized by the value of $\Omega_{GW}$) do
not exceed $10^{-8} - 10^{-9}$, although the examples of nonminimal variant of
the PBB scenario and some variants of preheating models (see Eq. (\ref{eq17}))
show that larger intensities ($10^{-7}$ or even $10^{-6}$) are not, in general, excluded.

The result, obtained in this paper, can be formulated in the following way:
the contribution to the total GWB, which is induced by primordial density fluctuations,
is constrained, in the region of $\sim 10^{-3} - 10^3$ Hz, as it is shown in Fig.
\ref{fig-gw-limit}.

This bound should be taken into account in elaborating inflationary models because, in
principle, the induced part of GWB (produced as a result of inflation) can significantly exceed the
first order part. The well known example is the running mass model \cite{Bugaev:2009zh}.

From the experimental point of view, the bound of Fig. \ref{fig-gw-limit} gives the maximum
possible signal from second order GWs. It is important because
if the experimentalists will be lucky do detect GWs above this bound, it will mean that
the signal is not due to induced GWB (e.g., nonminimal variants of the PBB scenario mentioned above
predict peak value of $\Omega_{GW}$ close to $10^{-6}$).

The bound shown in Fig. \ref{fig-gw-limit} is slightly more restrictive than the nucleosynthesis
bound \cite{Schwarztmann, Brustein:1996ut} (in LIGO region, at least). The latter bound depends on the
effective number of neutrino species $N_\nu$ while the former one is a straightforward
consequence ot the PBH limit on the spectrum of primordial density perturbations
(and Einstein's equations).

\paragraph*{Acknowledgments.}
We would like to thank Prof. A.A. Starobinsky for useful comments, and J.F. Dufaux
and R. Easther for drawing our attention to their papers on the subject.
The work was supported by FASI under state contract 02.740.11.5092.

\end{document}